\documentclass[proof]{WileyASNA-v1}

\newcommand*{\dtpartial}[1]{\dfrac{\partial#1}{\partial t}}
\newcommand*{\vv}{\boldsymbol{v}}

\articletype{Original Article}%

\received{8 September 2019}
\revised{XXX}
\accepted{XXX}

\begin{document}

\title{Metal abundances in the MACER simulations of the hot interstellar medium}

\author[1]{Silvia Pellegrini*}

\author[2,3]{Zhaoming Gan}

\author[3]{Jeremiah P. Ostriker}

\author[1]{Luca Ciotti}

\authormark{PELLEGRINI \textsc{et al}}

\address[1]{\orgdiv{Department of Physics and Astronomy}, \orgname{University of Bologna}, \orgaddress{\state{via Gobetti 93/2, 40129 Bologna}, \country{Italy}}}

\address[2]{\orgdiv{Shanghai Astronomical Observatory}, \orgname{Chinese Academy of Sciences}, \orgaddress{\state{80 Nandan Road, Shanghai}, \country{People's Republic of China}}}

\address[3]{\orgdiv{Department of Astronomy}, \orgname{Columbia University}, \orgaddress{\state{550 W, 120th Street, New York}, \country{NY 10027, USA}}}

\corres{*S. Pellegrini, Dept. of Physics and Astronomy, University of Bologna, via Gobetti 93/2, I-40129 Bologna. \email{silvia.pellegrini@unibo.it}}

\abstract{A hot plasma is the dominant phase of the interstellar medium of early-type galaxies. Its origin can reside in stellar mass 
losses, residual gas from the formation epoch, and accretion from
outside of the galaxies. Its evolution is linked to the dynamical 
structure of the host galaxy, to the supernova and AGN feedback, and to (late-epoch) star formation, in a way that has yet to
be fully undestood.  Important clues about the origin and evolution of the hot gas come from the abundances of heavy metals, 
that have been studied with increasing detail with XMM-Newton and
Chandra. We present recent high resolution hydrodynamical 
simulations of the hot gas evolution that include the above processes, and where several 
chemical species, originating in AGB stars and supernovae of type Ia and II, have also been considered.
The high resolution, of few parsecs in the central galactic region,
allows us to track the metal enrichment, transportation 
and dilution throughout the galaxy. The comparison of model results with observed abundances reveals a good
agreement for the region enriched by the AGN wind, but also discrepancies for the diffuse hot gas; the latter indicate the need for
a revision of standard  assumptions, and/or the importance of neglected effects as those due to the dust, and/or residual uncertainties 
in deriving abundances from the X-ray spectra.}


\jnlcitation{\cname{%
\author{Pellegrini S.}, 
\author{Z. Gan}, 
\author{J.P. Ostriker}, and 
\author{L. Ciotti}} 
(\cyear{2019}), 
\ctitle{Metal abundances in the MACER simulations of the hot interstellar medium}, 
\cjournal{Astron. Nachr.}, \cvol{2019;00:1--6}.}

\maketitle

\section{Introduction}\label{sec1}
A hot plasma dominates the ISM of early-type galaxies, as discovered with {\it Einstein} X-ray observations,
and investigated thoroughly until recently with {\it Chandra} and {\it XMM-Newton}~\citep{Fabbiano2012}.
The  X-ray emission of this hot phase correlates with the main galaxy
properties, as the optical luminosity, and the stellar and total galaxy
mass, although with a significant
scatter~\citep{Goulding2016,Forbes2017}. The observed X-ray properties have led to a picture where the
origin of the hot ISM is in the old, ageing stellar population, and perhaps also 
in circumgalactic infall~\citep{Pell2018,Werner2019}; its evolution
is the result of the main galactic properties, as the depth and shape of the potential well, and the
stellar kinematics, and of important phenomena, such as heating from supernovae and AGN feedback,
and late-epoch star formation [\cite{Cio2017,Gan18}, hereafter G19a; \cite{Negri2014}]. While the respective roles of these
processes are not fully understood yet, it is expected that they contribute to the enrichment of the gas in metals,
and, that, during its evolution,  they are able to circulate the metals within and
outside the galaxies. Thus, metals provide important diagnostic tests
for the ISM origin and the mentioned processes regulating its evolution.
In the hydrodynamical MACER code for the evolution of the ISM of massive elliptical
galaxies (G19a), element tracers have recently been added in order to track
the metal enrichment, transportation, and dilution  throughout the galaxy, and outside [\cite{Gan19}, hereafter G19b].
In the following we briefly describe the MACER modeling (Sect. 2), with particular emphasis on 
the sources of mass and metals for the hot ISM (Sect. 2.1); we then present the 
results of the simulations for the metals only (Sect. 3), and the conclusions  (Sect. 4) that include
a comparison with observations.

\section{Modeling the ISM evolution}\label{sec2}
The physical modeling of the ISM evolution has been implemented in
grid-type hydrodynamical simulations, performed with the Athena++
code (see G19a,b). The main advantages of these
simulations are the accurate treatment of the sources of mass and
heating, the latter from SNe and the AGN, and the high
resolution, that reaches 2.5 pc at the center, so that the fiducial
Bondi radius for accretion onto the central massive black hole
(hereafter MBH) is resolved. The ISM evolution is followed from an age
of 2 Gyr (after the main galaxy formation phase) until the present
epoch, for a time interval of 11 Gyr.

The code solves the time-dependent Eulerian hydrodynamical equations
with sources and sinks of mass, energy, and momentum. Sources of mass
are provided by three channels: 1) the original, ageing stellar
population, 2) the new stars [cold, dense gas is allowed to form stars, following both a Jeans criterion and Toomre
instability criterion; see G19a for more details], and 3) accretion from the circumgalactic
medium (hereafter CGM). No mass loading of cold gas into the hot phase is considered.
These mass sources in turn provide metals
(Sect. 2.1).  Sources of heating for the gas come from SNe explosions (of types SNIa and SNII, the latter only
from the new stars, with a thermalization efficiency of 0.85), from the thermalization of the stellar motions, and from accretion
onto the MBH.  Accretion produces a nuclear luminosity, with an
efficiency scaling with the mass accretion rate, with the consequent radiative
feedback; it also produces mechanical feedback via AGN winds, with maximum velocities
of $10^4$ km s$^{-1}$, and a mechanical efficiency again scaling  with the mass accretion rate,
as observed for nuclear outflows~\citep{Arav13,Carniani15,Tombesi15}. More details are given in G19a.

An important feature of the modeling is the implementation of
streaming velocity for the stars; this imparts net angular momentum to their
mass losses, so that the gas, when infalling in a cooling flow, settles in a rotationally supported 
disk-like structure; gas then cools in the disk, where it  becomes
gravitationally unstable, according to the Toomre Instability, and new stars form.
Angular momentum transfer and star formation are then allowed at the same time,
with the consequent near coincidence of star formation and AGN activity.

\begin{figure}
\vskip -4.5truecm 
\includegraphics[width=92mm,height=99mm]{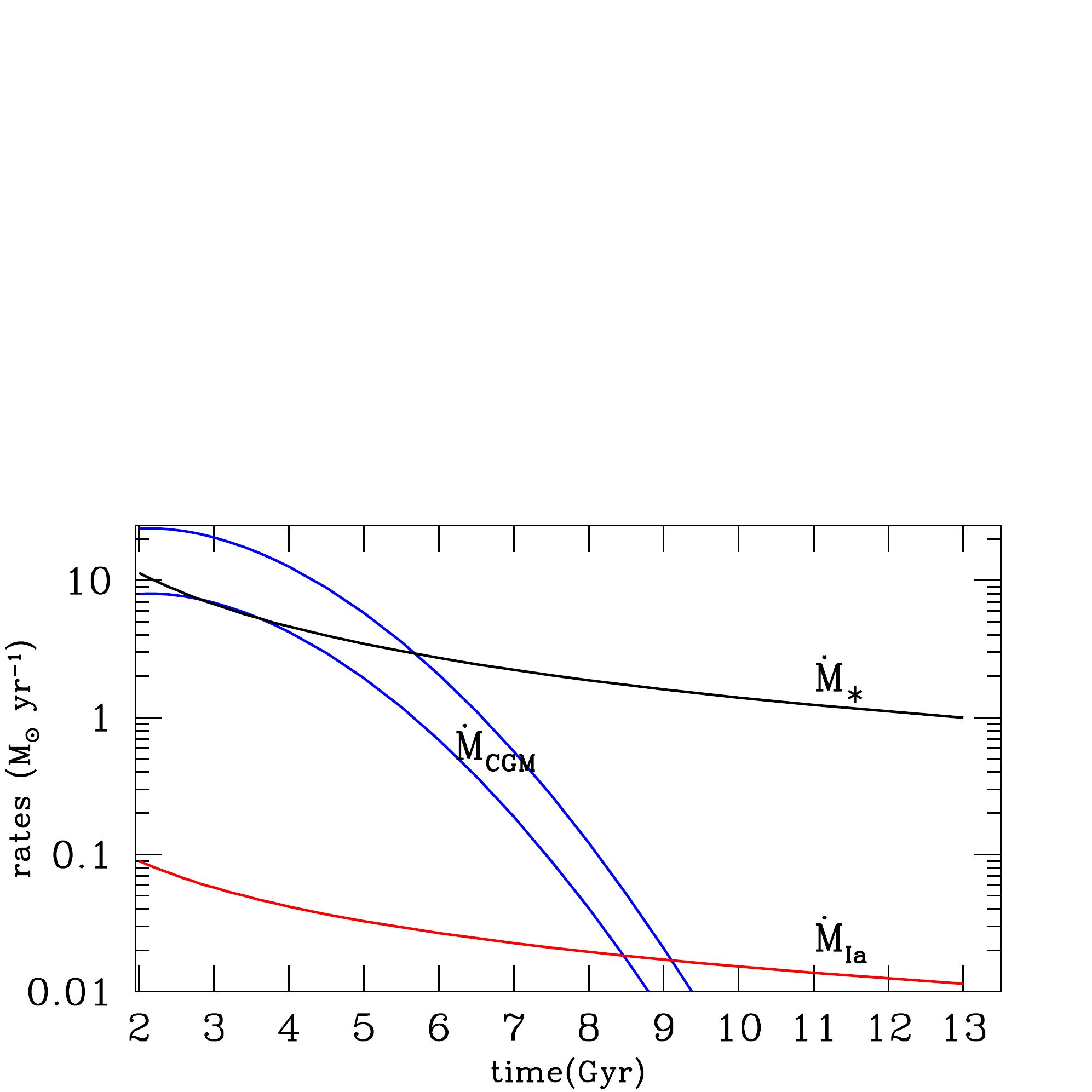}
\caption{The mass input rates to the ISM from the old stellar population (AGB winds, 
  in black, $\dot M_{\star}$; and SNIa's ejecta, in red,  $\dot M_{\rm{Ia}}$), 
  and from the cosmological CGM infall (in blue, $\dot M_{\rm{CGM}}$, for the two normalizations 
  of CGM$_{20}$ and CGM$_7$ models). 
  The $\dot M_{\star}$ and $\dot M_{\rm{Ia}}$ rates refer to the  galaxy with  $M_{\star}= 3.3\times  10^{11}$M$_{\odot}$
  of Sect.~\ref{results}.  See Sect.~\ref{sources} for more details.}
\label{fig1}
\end{figure}

\begin{figure*}[t]
  \vskip -1truecm
\centerline{\includegraphics[width=58mm,height=77mm]{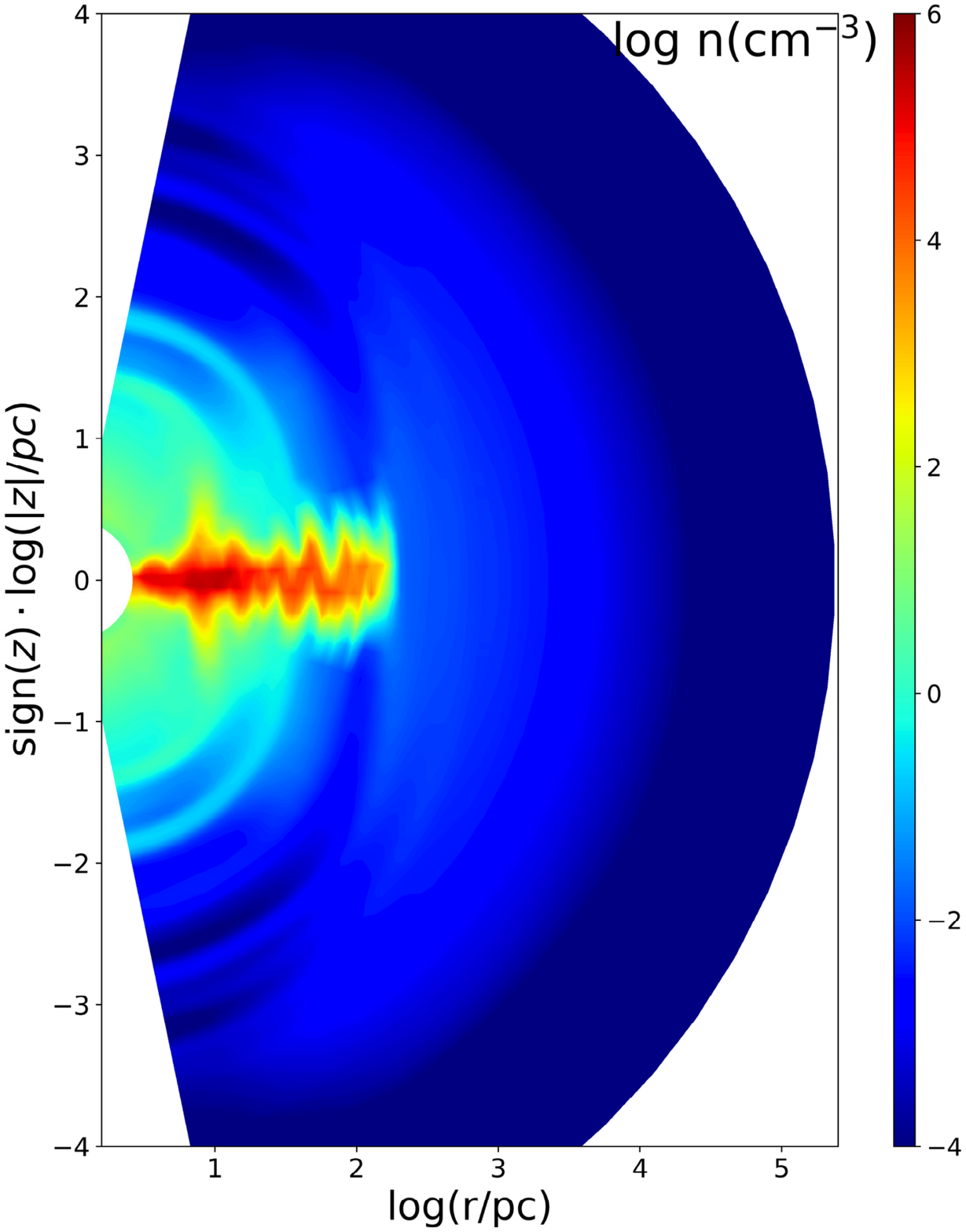}
  \includegraphics[width=58mm,height=77mm]{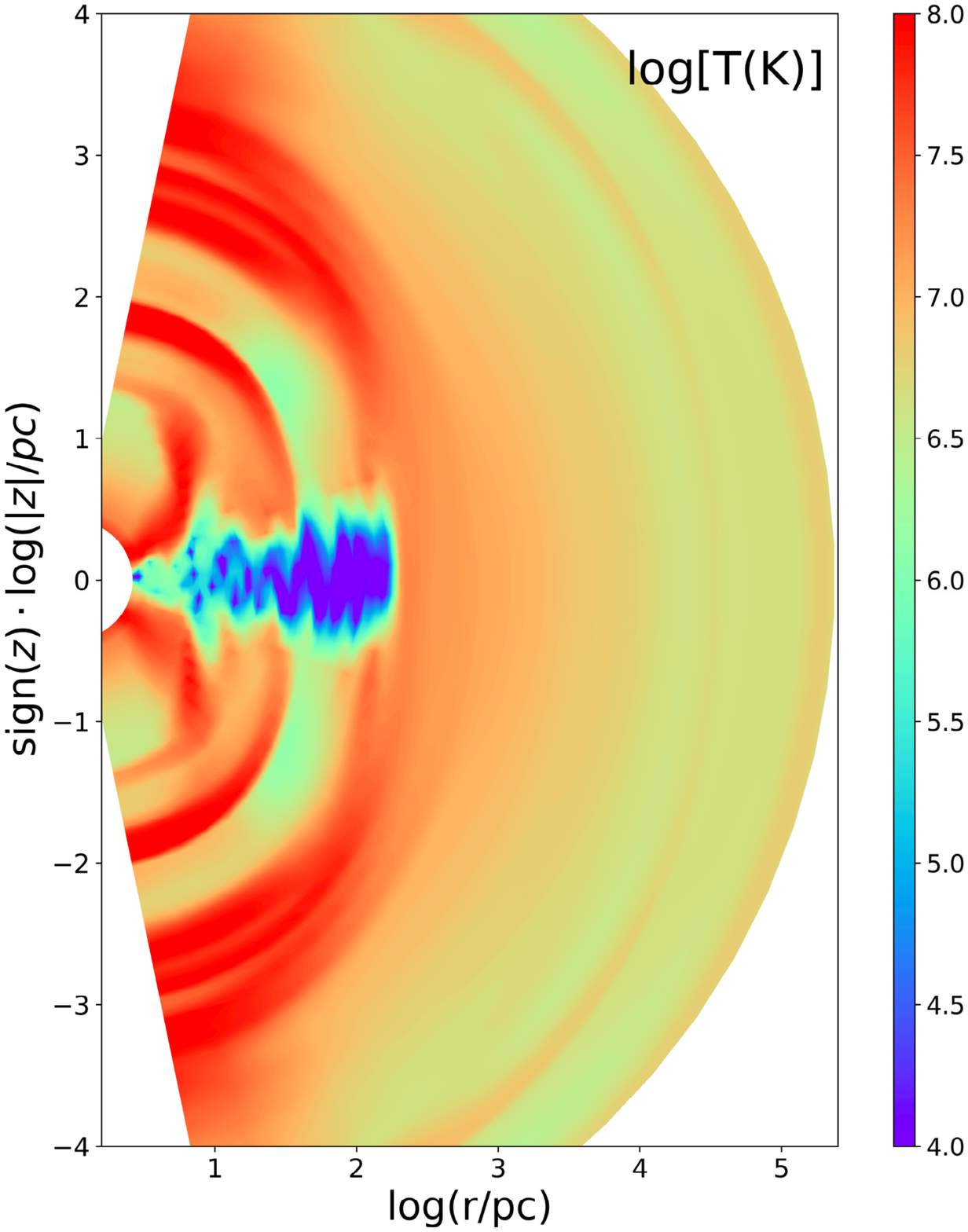}
  \includegraphics[width=58mm,height=77mm]{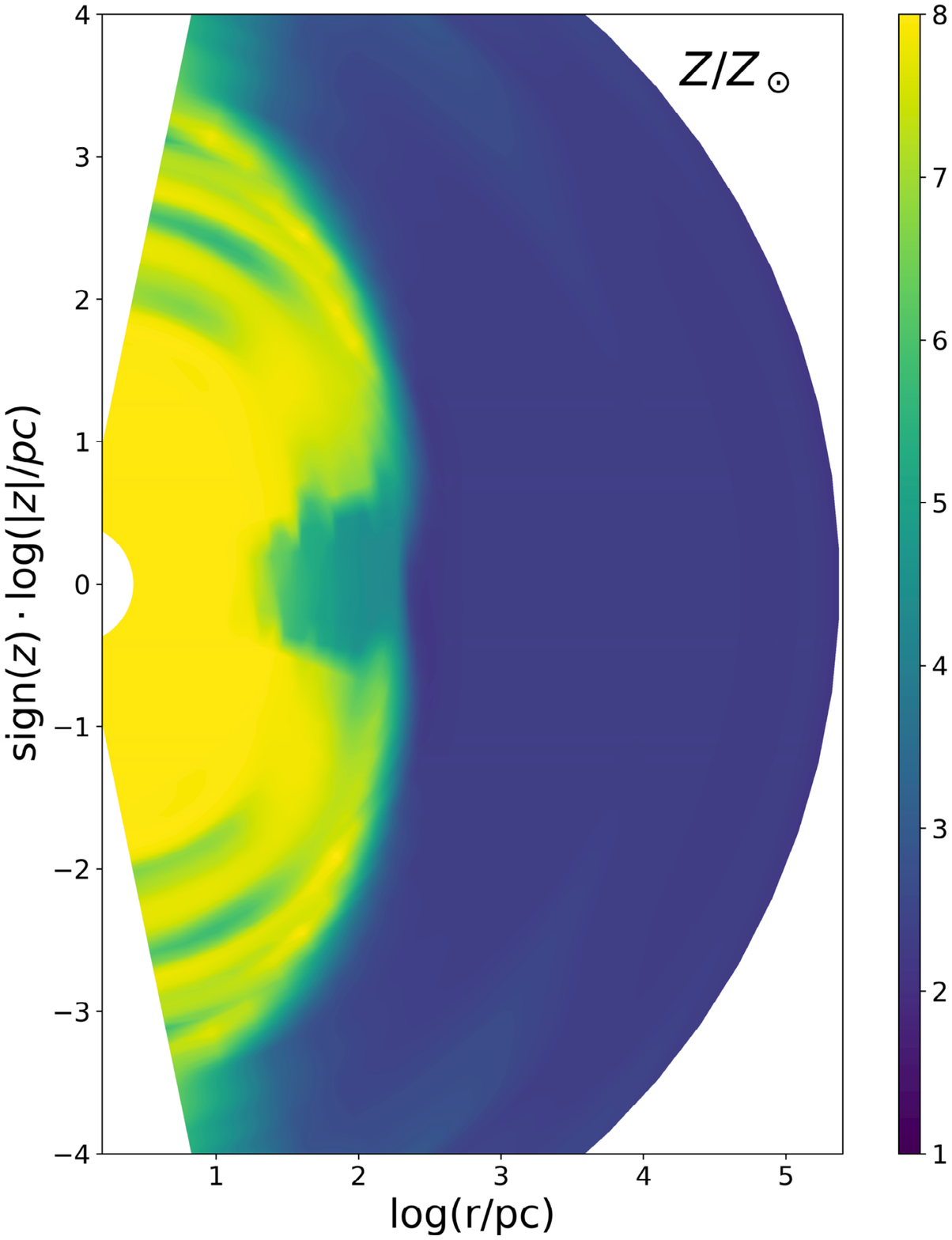}}
\vskip -0.5truecm
\caption{Meridional sections of the ISM density (left), 
  temperature (middle), and abundance Z of all metals (right), for the ISM in the galaxy model described in 
  Sect.~\ref{results}, during a nuclear outburst. The cold disk is evident, as the increase in Z
  above and below it, produced by the AGN wind.\label{fig2}}
\end{figure*}

\subsection{Sources of mass and metals}\label{sources}
Mass sources to the ISM are due to stars and the CGM.
The initial, old stellar population (hereafter SSP) injects mass via AGB
winds, at a mass rate $\dot M_{\star}$, and, to a much lesser extent, via SNIa's ejecta
at a mass rate $\dot M_{\rm{Ia}}$. As described in~\cite{Pellegrini2012}, the rates of these processes,
after an age $t_{\rm  age} =2$ Gyr for the SSP, are:
\begin{equation}
  \dot M_{\star}=10^{-12}A {M_{\star}\over M_{\odot}} \left( {t_{\rm
        age}\over 12{\rm Gyr}}\right)^{-1.3}\quad\quad M_{\odot} {\rm  yr}^{-1}
\end{equation}
where $A=3.3$ for the Kroupa IMF, and $M_{\star}$ is the SSP mass at $t_{\rm  age} =12$ Gyr, and:
\begin{equation}
  \dot M_{\rm Ia}=0.216\times 10^{-12} {L_{\rm B}\over L_{B\odot}}  \left( {t_{\rm
        age}\over 12{\rm Gyr}}\right)^{-1.1}\quad\quad M_{\odot} {\rm  yr}^{-1}
\end{equation}
where $L_{\rm B} $ is the present-epoch B-band luminosity of the SSP; and 1.35$M_{\odot}$ for the mass ejected in one SNIa event,
and a Hubble constant of 70 km s$^{-1}$ Mpc$^{-1}$, have been assumed.

New stars form in the cold gas disk with a top-heavy IMF of index equal to $-1.65$, based on 
observations (e.g., the central stellar disks in the Milky Way and
M31), and theoretical expectations~\citep{Bartko2010,Jiang2011,Lu2013}.
The chosen IMF has $\sim 60$\% of the new star mass in massive stars  which turn into SNII; for the new stars,
only SNII provide a source of mass and metals to the ISM (G19b). 

Infall of the CGM onto the galaxy also provides mass to the ISM, at a 
rate ($\dot M_{\rm{CGM}}$) estimated from cosmological zoom-in simulations, based on
the evolution of the accretion rate for 30 massive ellipticals
[\cite{Brennan2018}, G19a]. The rate declines sharply as expected in a $\Lambda-$CDM universe.
Figure~\ref{fig1} shows a comparison of the mass input rates from the SSP
and the CGM infall, during the time span of the simulations.
We adopt two normalizations for the CGM infall rate, within the range shown by
cosmological simulations; the corresponding total accreted mass during
the time interval in Figure~\ref{fig1} is 1/20
(for CGM$_{20}$ models) and 1/7 (for CGM$_7$ models) of $M_{\star}$.  For comparison, the total injected mass from the SSP
amounts to $M_{\star}/10$, during the same time.

The three sources of mass considered above (old and new stars, and the CGM) also provide metals to the ISM. In the code
the chemical evolution of the ISM due to the injection, dilution, and transportation of metals is followed
introducing 12 tracers $X_i$ ($i$=1, 2, ... 12), where $X_i$ is the mass of the $i$-th element per unit volume. The
elements considered are H, He, C, N, O, Ne, Mg, Si, S, Ca,
Fe, and Ni. Thus, 12 additional continuity equations are solved, assuming the
chemical species comove after they are injected into the ISM:
\begin{equation}
  \dtpartial{X_i} + \nabla \cdot (X_i\vv) + \nabla \cdot {\dot {\bf m}_{Q,i}} = \dot{X_i}_\mathrm{,\star} + \dot{X_i}_\mathrm{,Ia}  
  + \dot{X_i}_\mathrm{,II} - \dot{X_i}_\mathrm{,SF} 
\label{continuity},
\end{equation}
where $\vv$ is the gas velocity; the different metal-enriching sources from AGB winds $(\dot{X_i}_\mathrm{,\star})$,
SNIa's $(\dot{X_i}_\mathrm{,Ia}$), and SNII's ($ \dot{X_i}_\mathrm{,II}$) are evidenced; star formation $\dot{X_i}_\mathrm{,SF}=
(X_i/\rho) \cdot  {\dot {\rho}_\mathrm{SF}}$, where $\rho$ is the gas density and $ {\dot {\rho}_\mathrm{SF}}$ is the gas sink 
due to star formation, is treated as a sink of local metals; and
$  {\dot {\bf m}_{Q,i}} = (X_i/\rho) \cdot {\dot {\bf m}_Q}$, where $ {\dot {\bf m}_Q}$ is the momentum density for the gas that becomes
 Toomre-unstable  (see G19b for more details).

The simulations start at an age of 2 Gyr, therefore the SSP contributes metals only via AGB winds
and SNIa's ejecta. The weakly time-dependent metal input from AGB winds 
was calculated with “CELib,” an open-source software
library for chemical evolution~\citep{Saitoh2017}. The abundance of the SSP is 
Z$_\mathrm{\star}$=1.5Z$_{\odot}$ [where Z$_{\odot}=0.0134$,~\cite{Asplund2009}], as appropriate for a
massive elliptical as the modeled one~\citep{Thomas2010}.
The nucleosynthetic yields of SNIa's are those of the N100 model of~\cite{Seitenz2013},
 dominated by Fe and Si.
The SNII yields are those of~\cite{Nomoto2013}, and average
values are calculated for the progenitor's masses born with the adopted IMF of the new stars
(see G19b for more details).

The CGM infall provides low-metallicity gas with Z$_{\rm CGM}=0.002$ (i.e., Z$_\mathrm{CGM}$=0.15Z$_{\odot}$).
This gas accretes through the galaxy outskirts, and produces a dilution of the abundance of the ISM of internal origin
as it falls into the galaxy and mixes with the ISM.
It does not appear explicitly in  equation~\ref{continuity}, since the CGM infall is implemented as an outer boundary
condition.

\begin{figure*}[t]
 \vskip -2.5truecm
\centerline{
  \includegraphics[width=89mm,height=124mm]{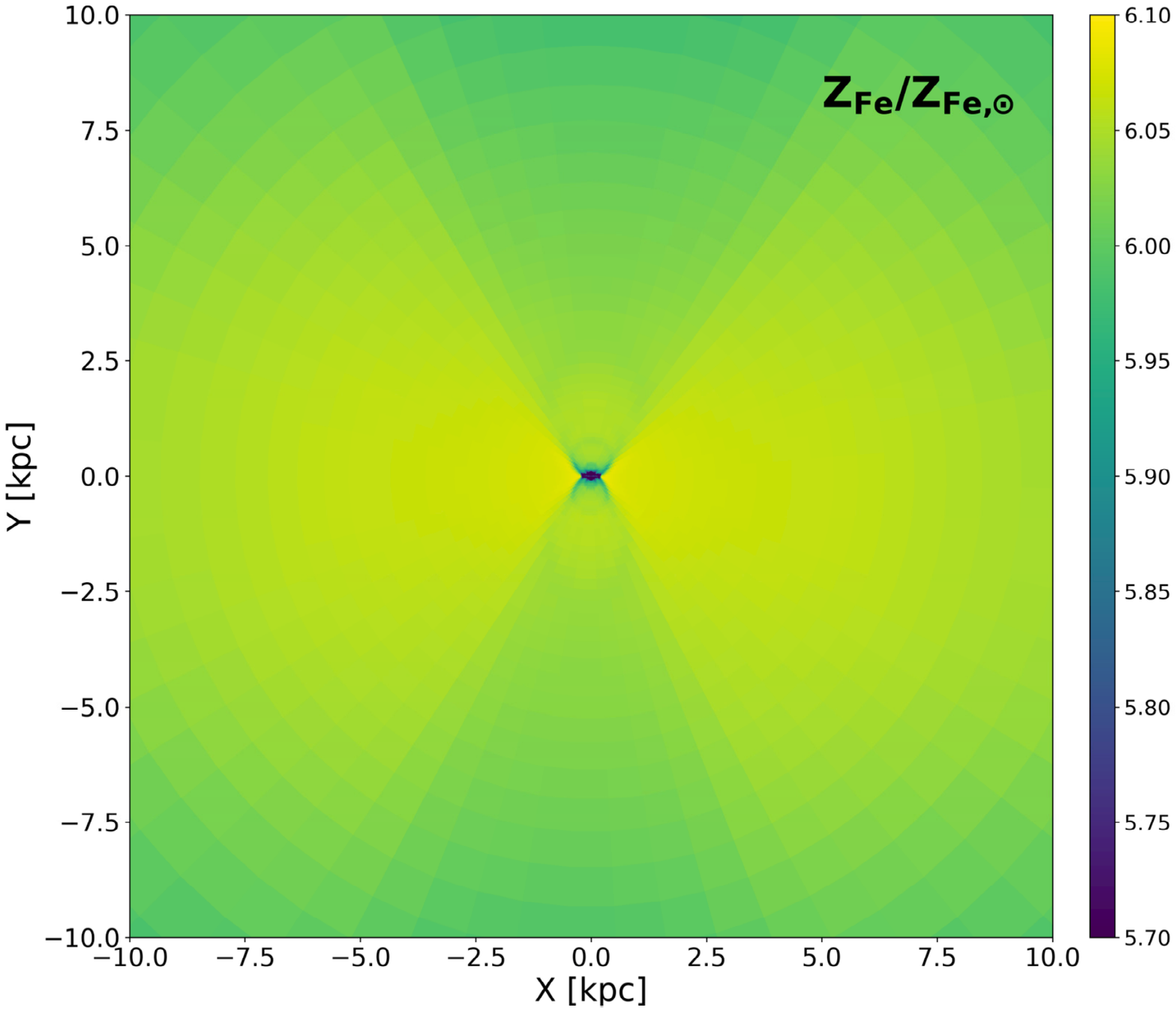}
  \includegraphics[width=89mm,height=124mm]{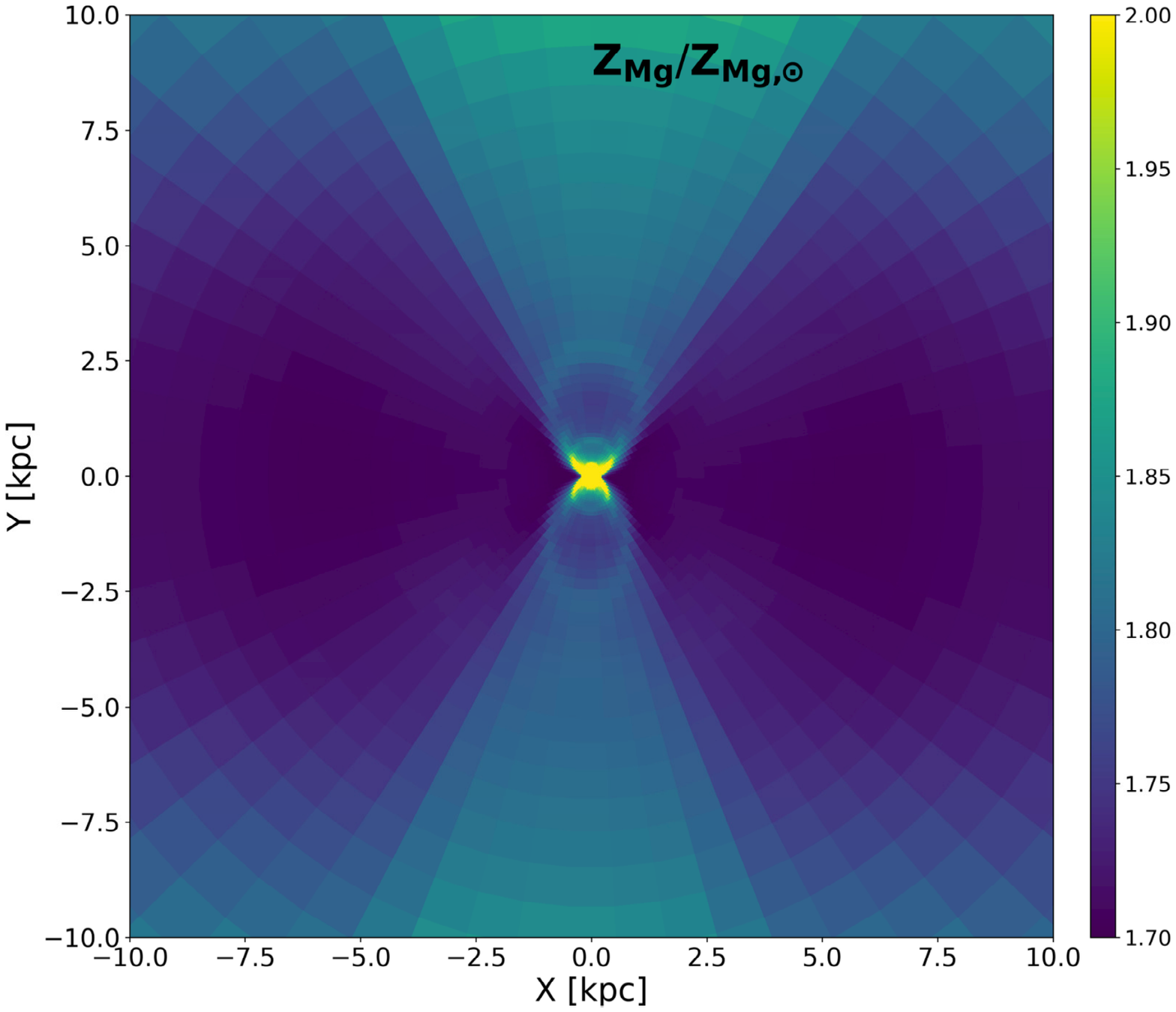}
}
\vskip -2truecm
\caption{Map of the emission weighted and projected abundance of Fe (left)
and Mg (right),  for the ISM in the galaxy model described in
Sect.~\ref{results}; the snapshot is taken during a nuclear outburst, at the same time of Figure~\ref{fig2};
note the linear scale instead of the logarithmic one used in Figure~\ref{fig2}.
\label{fig3}}
\end{figure*}

\section{Results}\label{results}
We present the resulting chemical evolution of the ISM for an E2 elliptical galaxy, of stellar mass at the
present epoch of $M_{\star}= 3.3\times  10^{11}$M$_{\odot}$, effective radius
$R_e=7$ kpc, and central stellar velocity dispersion of 280 km$s^{-1}$. A central MBH and a spherical dark matter halo
are also added, and the ordered rotational velocity of the
stellar component is given by the Satoh decomposition, with exponentially declining rotation parameter $k$
 (see G19b). The presence of streaming velocity results for the ISM into the formation  of
a central cold disk, where star formation takes place; the latter becomes then an important
metal-enriching source, mostly by SNe II. In fact, the ISM evolution undergoes recurrent cycles
during which the gas accumulates in the disk, until when the Toomre instability sets in, allowing for
star formation and mass inflow from the disk towards the MBH. Thus, with a short delay,
star formation is followed by accretion of disk material onto the MBH, whose feedback action 
via biconical BAL (broad absorption line) winds causes  the ejection of  recently metal-enriched gas from the disk (Figure~\ref{fig2}).
The abundance of the BAL winds is evaluated by
tracking the chemical composition of the mass accreted via the inner boundary; 
the metals  passing through the center are present in the BAL winds and 
injected  back to the galaxy from its nucleus. Freshly produced metals are
spread  out to large distances (of $\simeq 1$ kpc and more); the abundance of the high velocity, outward-moving gas,
can be many times solar.
Note that these high abundances are fundamental for the gas
surrounding the MBH to be multiphase, and include a stable gas phase
representing the broad line region (BLR) clouds~\citep{Chakravorty09}.

Figure~\ref{fig2} shows the density and temperature of the ISM during
an outburst. The cold disk, extending for $r\simeq 150$ pc, is clearly visible, as is the biconical wind structure 
in the temperature map. 
Note that the size of the cold disk strongly depends on the galaxy rotation profile. In the right panel the
total abundance Z is shown (the logarithmic scale on the axes emphasizes the central region).
SNII products, transported by the winds, contaminate the region above  and below the disk; they reveal 
many subsequent ejection episodes closely spaced in time,  that took place during the nuclear outburst.
As an example,  in the outflowing region affected by  the BAL wind with a radial velocity $v_r \geq 10^3$ km s$^{-1}$,
the mass-weighted average abundance is $4 Z_{\odot}$ (with a weighting towards products from SNII).
  
Figure~\ref{fig3} shows the emission weighted and projected maps for the abundance of Fe and Mg, at the same time 
of Figure~\ref{fig2} during the outburst.
Interestingly, these two metal species show an opposite behavior. In fact, star formation increases the overall Z
of the ISM, but  in the SNII ejecta the Fe is less abundant than in the mass lost by the old SSP.
This causes Fe to be more abundant over the main galaxy body, and less abundant in the region affected by recent star formation
(the central disk and the biconical region affected by the spreading of metals due to the BAL winds).
The symmetric behavior is shown by Mg, that is more abundant in the regions affected by SNII ejecta.
Thus, SNII-products as Mg are a tracer of the
distances reached by the ISM driven by the AGN winds. Note that the biconical region in Figure~\ref{fig2} comes from
the subsequent action of many outbursts during the galaxy's lifetime, and that each outburst produces an abundance
pattern rich in substructure, that cannot be shown here due to the limited space.

\begin{figure}[t]
  \vskip -2truecm 
\includegraphics[width=89mm,height=80mm]{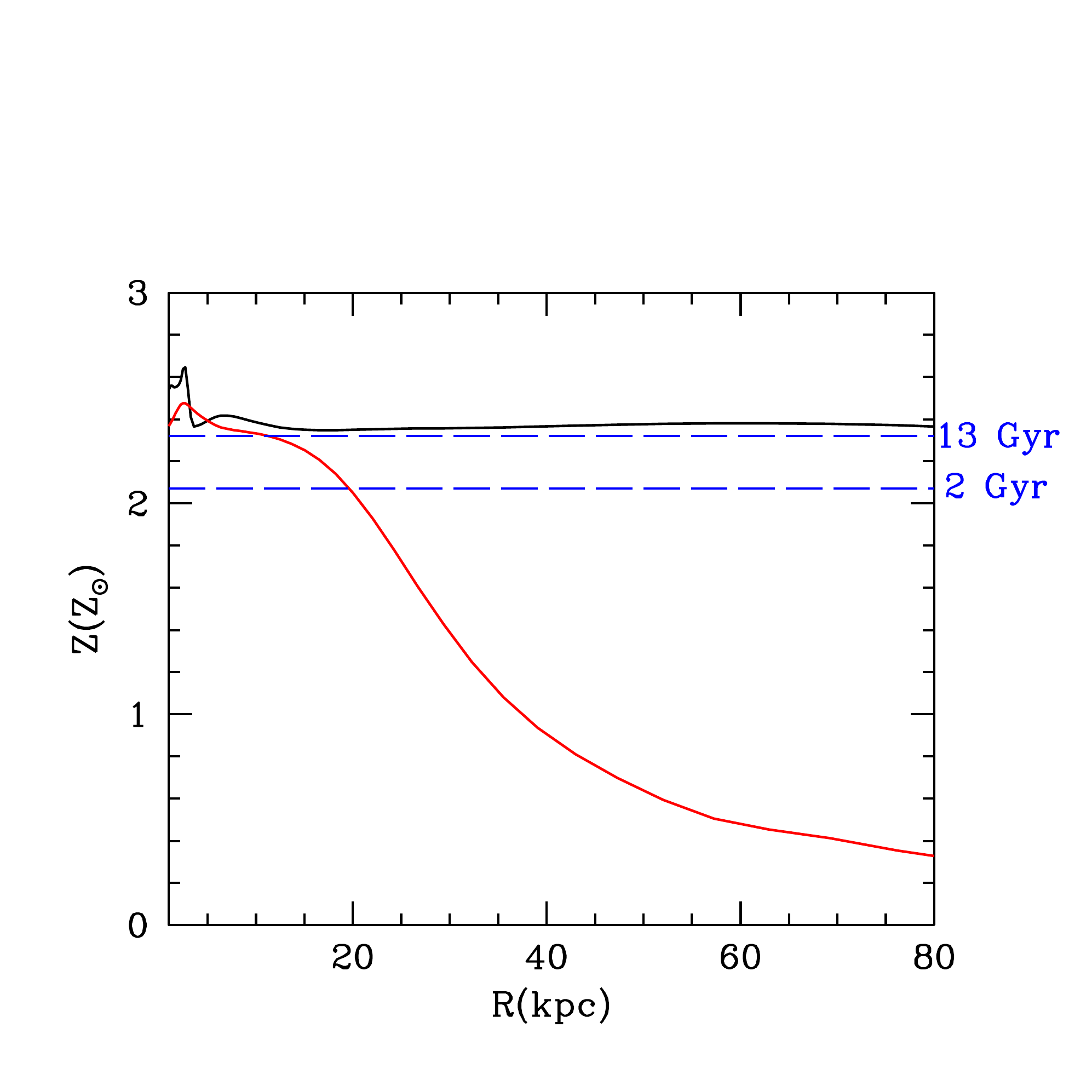}
  \vskip -0.5truecm 
\caption{Emission-weighted, projected and circularized abundance of the ISM Z(R), 
  for the  CGM$_{20}$ (in black) and  CGM$_{7}$ (in red) models, for the galaxy described in Sect.~\ref{results}, 
  at an age of 12 Gyr. 
Also shown (in blue) are the abundances Z$_{old}$ at two epochs, for the mass ejected 
by AGB winds+SNIa's (with Z$_{\star}=1.5Z_{\odot}$ for the SSP). The Z$_{old}$ values are independent of the 
galaxy stellar mass, provided that the same IMF and Z$_{\star}$ apply  (see Sect.~\ref{results} for more details).}
\label{fig4}
\end{figure}

Figure~\ref{fig4} shows the emission-weighted, projected and circularized abundance profile, Z(R), for the ISM when the
SSP age is 12 Gyr. Also shown is the  abundance of the  mass contributed by the SSP sources (AGB+SNIa's; hereafter
``passive'' abundance Z$_{old}$). Since the  SSP is assumed to be homogeneous across
the galaxy, Z$_{old}$  is position-independent; it depends instead on the SSP
age, since the metal injection of AGB winds and of SNIa's follows different time-dependencies (Sect.~\ref{sources}).
In Figure~\ref{fig4} the values of Z$_{old}$ at two ages of the SSP (2 and 13 Gyr) are shown. One notices that 1) Z$_{old}$  varies in a narrow
range\footnote{The ``passive'' values in Figure~\ref{fig4} refer to
  the Kroupa IMF, but  they would change very little for the Salpeter IMF.}, from  2.1
to 2.3 $Z_{\odot}$, over the whole time-span, and 2)  Z(R) of the ISM
when the SSP  is 12 Gyr old (black line) is
close to Z$_{old}(13$ Gyr): Z(R) is just slighly above it, due to the
enrichment contributed by new stars, and does not show pronounced radial trends.
The black line in the figure refers to the CGM$_{20}$ model, the same
shown in Figures 2 and 3; the red line refers instead to its CGM$_7$
version, with a larger CGM infall (Sect. 2.1). The CGM$_7$ model shows a large radial gradient:
its Z(R) is comparable to that of  CGM$_{20}$  in the central galactic region, but decreases  outer of $R\sim 15 $ kpc.
The abundance of the ISM averaged {\it over the whole galaxy} is 2.4 $Z_{\odot}$ for the  CGM$_{20}$ model,
dropping to $\sim 1.5 Z_{\odot}$ for the CGM$_{7}$ one.

\begin{figure}[t]
  \vskip -2truecm 
\includegraphics[width=89mm,height=80mm]{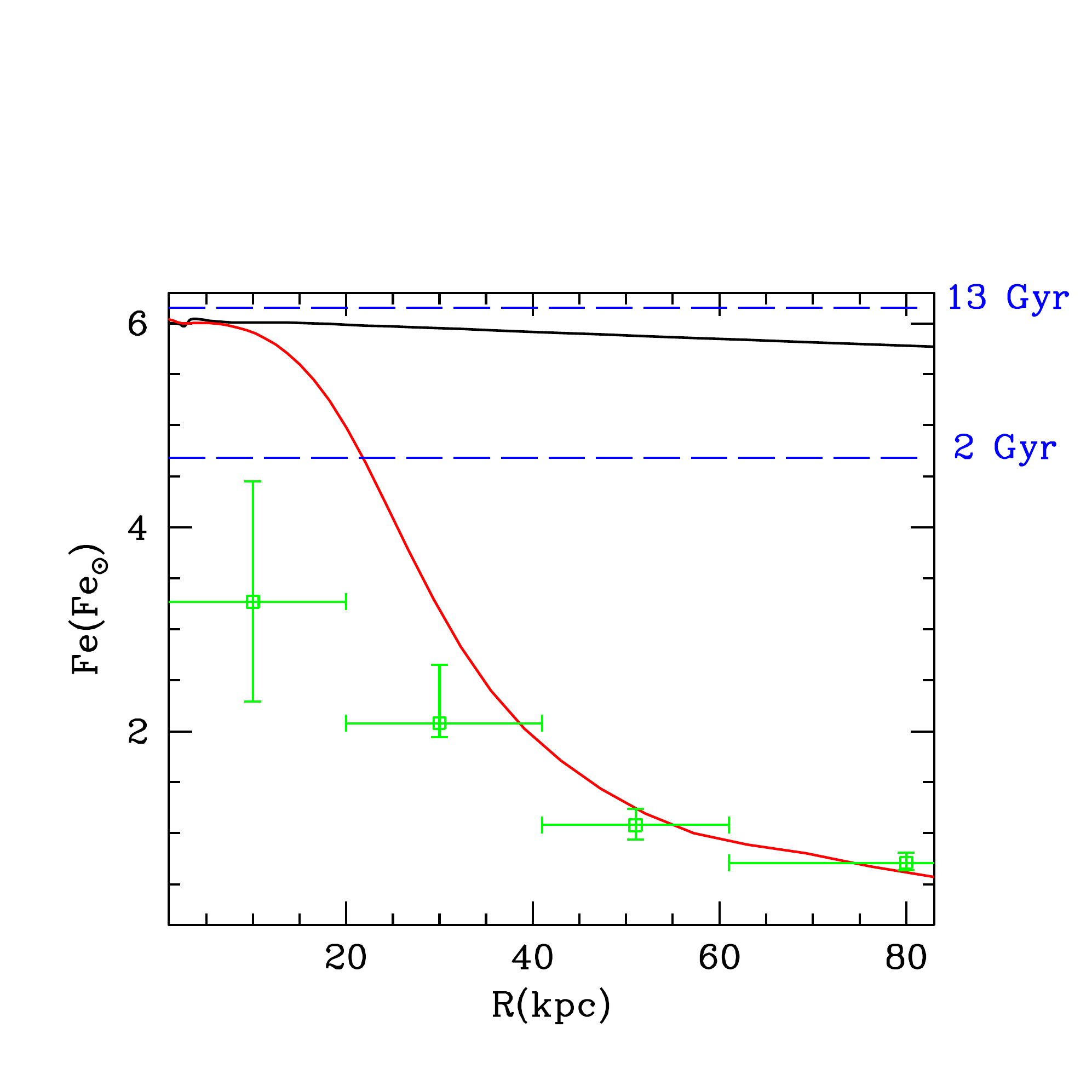}
  \vskip -0.5truecm 
\caption{Emission-weighted, projected and circularized Fe abundance of the ISM, Fe(R), 
  for the  CGM$_{20}$ (in black) and  CGM$_{7}$ (in red) models, for the galaxy described in Sect.~\ref{results}, 
  at an age of 12 Gyr. 
Also shown (in blue) is Fe$_{old}$ at two epochs, for the mass ejected 
by AGB winds+SNIa's (with Fe$_{\star}=1.5Fe_{\odot}$ for the SSP). Green squares show  Fe(R) for NGC507, a dominant
elliptical galaxy in a small group, from XMM-Newton observations~\citep{KF04}. See Sect.~\ref{results} for more details.}
\label{fig5}
\end{figure}

Figure~\ref{fig5} shows the emission-weighted, projected and
circularized Fe abundance profile, Fe(R), for the ISM when the SSP age is 12 Gyr, again with the ``passive'' Fe abundance (Fe$_{old}$)
at 2 and 13 Gyr of age for the SSP, and the CGM$_7$ model included for
comparison. The same trends of 
Figure~\ref{fig4} are apparent, with one difference: Fe(R) keeps below Fe$_{old}$(13 Gyr), even at
the center. This is explained by the fact that Fe$_{old}$ is time-increasing, and the SSP age for the plotted ISM is 12 Gyr;
and by the large difference in the Fe-production between the SPP
and the new stars (see the discussion above for Figure~\ref{fig3}).
Figure~\ref{fig5} also shows the observed Fe(R) for a giant elliptical galaxy: except for the innermost point, outer of
R$\approx$20 kpc the agreement with the modeled ISM is good.

\section{Conclusions}
In the MACER code for the investigation of the evolution of the ISM of elliptical galaxies, at high
spatial resolution and with AGN feedback, the metal enrichment, transportation, and dilution
processes have recently been implemented (G19b). From these recent simulations for a massive
elliptical, we have here emphasized the following results:
\begin{itemize}
\item the average Z of the hot ISM on the whole galactic scale is $\simeq 2.4Z_{\odot}$; this value can go
  down to $\sim 1.5Z_{\odot}$ when the CGM infall rate is closer to the largest estimates from cosmological simulations.
\item for  R < 15 kpc,  Z(R) of the hot ISM is close to the 'passive' prediction (i.e., the abundance of AGB winds+SNIa's),
  independent of the cosmologically-motivated CGM infall rate. A small
  increase due to new star formation is present; this increase is  disk-dependent,
  and thus related to the streaming velocity field of the host galaxy.
  Outer of $R\simeq 15$ kpc, Z(R) sharply decreases for the larger CGM infall rate.
 \item the Fe abundance profile Fe(R) follows the same trends shown by Z(R), with the difference of being {\it lowered} by the
  presence of new star formation. This feature is also shown in the abundance map of Fe, and is opposite to that shown in
  the Mg abundance map.
\item during outbursts, the outflowing regions, driven by the BAL wind, can reach an abundance of Z$\simeq 4Z_{\odot}$, and
  can transport metals (mostly SNII products made in the disk) out to large radii. This finding is in agreement 
  with observations that Z of a quasar BLR in massive galaxies is  0.3 to 1.0 dex larger than Z of the host galaxies, and does
  not evolve with cosmic time, which implies that the  metal enrichment  is due to recent star formation, as in the
  simulations~\citep{Nagao06, Xu2018}.
\end{itemize}
While the results above are encouraging for the part concerning the role of AGN winds, the observed Z (and Fe) of the hot ISM
reveal more problematic issues. Observed values are generally solar, or slighlty supersolar, except for a few cases (one is
shown in Figure~\ref{fig5}), as established even recently, with the improved quality of both $Chandra$ and XMM-$Newton$
observations and data analysis~\citep{Hum2006,Loew2010,Mernier17,Mernier18}. The observed Z, and especially Fe, are lower
than predicted by models, which represents an old problem, that has been alleviated in more
recent investigations, but is clearly persisting. 
Contributing factors to it include residual uncertainties in the derived abundances, as for example due to
the degeneracy of metallicity and emission measure, and the multi-temperature thermal structure of the ISM~\citep{Kim2012,Plaa2017}; 
dust depletion of metals~\citep{Panagoulia2013,Lakhchaura19}; and incomplete mixing of SNIa's
ejecta~\citep{Matsushita2000,Tang2010}. 
Based on the present results, the CGM infall can also have an important effect.
Another possibility is the mixing and stirring of metals between the inner and outer parts of the galaxies, as would be provided by
a larger feedback wind efficiency, and/or infall of  satellites.
Finally, we emphasize that the depletion of metals onto dust has not been inserted in the models yet, and thus 
elemental abundances correspond to the total (dust plus gas
phase) mass in that element. The depletion of refractory elements onto dust grains may
be a large correction~\citep{Hensley2014}, and this subject is reserved to a future work.

\nocite{*}
\bibliography{Wiley-ASNA}%

\end{document}